\documentclass[]{article}
\usepackage[T2A]{fontenc}
\usepackage[utf8]{inputenc} 
\pdfoutput=1
\usepackage[russian,english]{babel}
\usepackage{amsfonts,amssymb,amsmath}

\def\dac{\displaystyle\frac}

\def\d{\partial}
\def\[{\left[}
\def\]{\right]}
\def\({\left(}
\def\){\right)}

\def\be{\begin{equation}}
\def\ee{\end{equation}}

\begin{document}

\title{
Wave-like exact models with symmetry of spatial homogeneity in the quadratic theory of gravity with a scalar field
}


\author{K.E.~Osetrin\thanks{osetrin@tspu.edu.ru}, \\I.V.~Kirnos, \\E.K.~Osetrin, \\A.E.~Filippov\\
\it Tomsk State Pedagogical University}

\date{May 28, 2021}

\maketitle{}

\begin{abstract}
Exact solutions are obtained in the quadratic theory of gravity with a scalar field for wave-like models of space-time with spatial homogeneity symmetry and allowing the integration of the equations of motion of test particles in the Hamilton-Jacobi formalism by the method of separation of variables with separation of wave variables (Shapovalov spaces of type~{II}). The form of the scalar field and the scalar field functions included in the Lagrangian of the theory is found. The obtained exact solutions can describe the primary gravitational wave disturbances in the Universe (primary gravitational waves).
\end{abstract}

\section*{Introduction}

The task of the work is to construct exact models of primary gravitational wave disturbances of the Universe on the basis of quadratic theories of gravity with a scalar field in space-time models with symmetry of spatial homogeneity.
Curvature-quadratic scalar-field theories of gravity usually arise as consequences in models of quantum gravity, string theory, M-theories as modifications of general relativity (GR) \cite{Boulware1985}.
They have a number of interesting properties (renormalizability, inflation) and can claim a more realistic description of the initial stages of the development of the Universe in comparison with general relativity~\cite{Mandal2021},
\cite{Fomin2020}.
It should be noted that historically the first inflationary model
\cite{Starobinskii1979}
was built on the basis of the theory of gravity, quadratic in curvature, but models with a scalar field - an "inflaton" (whose development began after the work 
\cite{Guth1981347}.
For this reason, the properties of the inflaton are sufficiently well studied and it seems unjustified to abandon it, so we include the terms describing the scalar field in the Lagrangian along with the terms that are quadratic in curvature.

A review of theories of this kind and their comparison with cosmographic tests are given in the work
\cite{Odintsov2012155}.
Also, a large survey of theories with Lagrangians, nonlinear in curvature and containing the Gauss-Bonnet invariant, is given in the papers
\cite{Odintsov201159}, 
\cite{BambaOdintsov2015Symmetry220}, 
a number of solutions were obtained in works
\cite{Ivashchuk2015},
\cite{Ivashchuk2016},
\cite{Camci2018},
\cite{Bajardi2020},
\cite{Herdeiro2021}.
Theories with Lagrangians of the form $ f (\varphi, R) $ and their relationship with inflation and modern accelerated expansion are considered in the work
\cite{Odintsov20171}.

To describe the early stages of the development of the Universe, spatially homogeneous non-isotropic space-time models are considered more realistic models than the homogeneous and isotropic Friedman-Robertson-Walker models, especially since in many models
\cite{Heckmann1989},
\cite{Khalatnikov2003}
anisotropy decreases with time, space becomes isotropic
\cite{Ellis1997}.

The description of primary gravitational perturbations (primary gravitational waves), the occurrence of which is predicted at the early (quantum) stages of the development of the Universe, requires the use of adequate mathematical methods to describe such wave models of space-time.
It should be noted that not in all theories of this kind, the speed of a gravitational wave and the speed of light in a vacuum coincide. Meanwhile, the gravitational wave burst GW170817, associated with the merger of neutron stars, was accompanied by electromagnetic radiation, and the burst of electromagnetic radiation came a little later, which is associated with the physics of the process and, in addition, with the delay of electromagnetic radiation by the interstellar medium. Thus, we can assume that in a vacuum the velocities of gravitational and electromagnetic waves coincide. The question of the resulting restrictions on the form of the Lagrangian of a theory quadratic in curvature is studied in the papers
\cite{Jana2021},
\cite{Odintsov2021Annals},
\cite{Oikonomou2020}.

As adequate mathematical models for the considered physical problems, we propose the use of spatially homogeneous wave-like Shapovalov spaces  \cite{Osetrin2020Symmetry}.
These spaces allow the existence of ''privileged'' coordinate systems, where complete separation of variables in the equations of motion of test particles in the Hamilton-Jacobi formalism with separation of wave variables is possible, on which the metric of space-time explicitly depends in privileged coordinate systems (wave-like spaces).

\section{Quadratic gravity with scalar field}

Consider a quadratic theory of gravity with a scalar field $ \varphi $, the Lagrangian of which has the form (see f.e.\cite{Mandal2021}):
\be
\mathcal{L}=\dac{R+\gamma(\varphi) R^2}{2\varkappa^2}-\xi(\varphi)\mathcal{G}-\dac12\omega g^{\mu\nu}\d_\mu\varphi\d_\nu\varphi-V(\varphi)+\mathcal{L}_{\mbox{Matter}},
\label{Lagrangian}
\ee
where $ R $ is the scalar curvature, $ \varkappa $ is the gravitational constant,
$ \gamma (\varphi) $ and $ \xi (\varphi) $ are scalar field functions,
\mbox {$ V (\varphi) $ is the potential} of the scalar field, $ \mathcal{G} $ is the Gauss-Bonnet term, $ \omega $ is a constant,
$ \mathcal{L}_{\mbox{Matter}} $ - Lagrangian of material fields.

Quadratic theory of gravity with Lagrangian (\ref{Lagrangian}) gives us the following field equations:
$$
Q_{\mu\nu}=
\dac{1}{\varkappa^2}\[(1+2{\gamma}{} R)R_{\mu\nu}+2\(g_{\mu\nu}\Box\({\gamma}{} R\)-\nabla_\mu\nabla_\nu\({\gamma}{} R\)\)
-\dac{R+{\gamma} R^2}{2}g_{\mu\nu}
\]
$$
$$
\mbox{}
 -8\nabla_{\gamma}\nabla_\beta\[\xi{}R_{\mu\phantom{{\gamma}}\nu}^{\phantom{\mu}{\gamma}\phantom{\nu}\beta}\]
 -8\nabla_{\gamma}\nabla_\mu\[\xi{}{R^{\gamma}}_\nu\]
-8\nabla_{\gamma}\nabla_\nu\[\xi{}{R^{\gamma}}_\mu\] 
$$
$$
\mbox{}
+8 g_{\mu\nu} \nabla_{\gamma}\nabla_\beta\[\xi{}R^{{\gamma}\beta}\]
 +4\nabla_\mu\nabla_\nu\[\xi{}R\]
 -4\Box\[\xi{}R\]g_{\mu\nu}
+8\Box\[\xi{}R_{\mu\nu}\]
$$
$$
\mbox{}
+4\xi{}
\left(
 4 R_{\mu{\gamma}}{R_\nu}^{\gamma}-RR_{\mu\nu}
 -R_{\mu{\alpha}\beta\gamma}{R_\nu}^{{\alpha}\beta\gamma}
 \right)
$$
\be
\mbox{}
+
(\omega/2)\,
g_{\mu\nu} g^{{\gamma}\beta}\d_{\gamma}\varphi\d_\beta\varphi
-\omega{}\d_\mu\varphi\d_\nu\varphi
+
g_{\mu\nu}
\left(
V{}+\xi{}\mathcal{G}
\right)
=T_{\mu\nu}
,
\label{EqsField}
\ee
and the equation for the scalar field $ \varphi $ takes the form
\be
\omega{}\Box\varphi
+ \dac{R^2}{2\varkappa^2}{\gamma}'(\varphi) -\mathcal{G}\xi'(\varphi) -V'(\varphi)=0
,
\label{EqScalar}
\ee
where $ \nabla $ is the covariant derivative, $ \Box $ is the d'Alembert operator, the prime means the derivative with respect to the scalar field $ \varphi $.

\section{Wave-like Shapovalov spaces}

Wave-like Shapovalov spaces
\cite{Osetrin2020Symmetry}
arise as a subset of the {S}t{\"{a}}ckel spaces
\cite{Shapovalov1978I, Shapovalov1978II, Shapovalov1979},
admitting the existence of privileged coordinate systems, where the equation of motion of test particles in the Hamilton-Jacobi formalism
\be
g^{ij} S_{,i} S_{,j} = m^2
\ee
can be integrated in quadratures by the method of complete separation of variables. Here $ g^{ij} $ is the space-time metric, $ S $ is the test particle action function, $ m $ is the test particle mass.

Shapovalov spaces, by definition~\cite{Osetrin2020Symmetry}, allow separation of non-ignored 
''wave'' variables (on which the metric depends) along which the space-time interval is equal to zero. The ability to separate the wave variables on which the metric depends allows us to consider these space-time models as “wave-like”.

Note that since the motion of test particles in space-time is carried out along geodesic lines, then the {S}t{\"{a}}ckel and Shapovalov spaces admit integration in quadratures of the geodesic equations. The ability to accurately integrate the equations of motion of test particles makes {S}t{\"{a}}ckel spaces an important tool in gravity problems, including modified theories.
We also investigated both the general properties of {S}t{\"{a}}ckel spaces with dust matter \cite{Osetrin2016MPLA} and radiation \cite{Osetrin2017JMP}, and spatially homogeneous models of {S}t{\"{a}}ckel spaces that are of interest for cosmology \cite{Obukhov2002RPJ42},
\cite{Osetrin2006JPA6641}, \cite{Osetrin2019RPJ292}, \cite{Osetrin2020RPJ403}, \cite{Osetrin2020RPJ410}.

Recently, a number of new results have been obtained on the classes of spaces that allow integration by the method of separation of variables in the Hamilton-Jacobi equation for charged test particles and in the Klein-Gordon-Fock equation for scalar and electromagnetic fields \cite{Obukhov2020Symmetry}, \cite{Obukhov2020RPJ1126},   \cite{Obukhov2020IJGMMP}, \cite{Obukhov2021IJGMMP}, \cite{Obukhov2021Symmetry}.

On the basis of the Steckel and Shapovalov spaces, a number of exact models have been built both for the general theory of relativity
(f.e.~\cite{Bagrov1988GRG1141},
\cite{Bagrov1997RPJ995}
)
and for modified theories of gravity
(f.e.~\cite{Osetrin2018RPJ1383},
\cite{Osetrin2020IJGMMP}, \cite{Osetrin2020MPLA}
).

In four-dimensional space-time, there are three classes of Shapovalov spaces according to the dimension of the Abelian group of motions they admit \cite{Osetrin2020Symmetry}. Type I Shapovalov spaces admit one Killing vector and the metric of this space in a privileged coordinate system depends on three ''non-ignored'' variables, including the ''wave '' variable, type II spaces admit two commuting Killing vectors and its metric in a privileged coordinate system depends on two “non-ignored” variables, including the “wave” variable and, finally, the metric of the type III Shapovalov space in the privileged coordinate system depends on only one wave variable.

In this paper, we will consider Type II Shapovalov spacetimes admitting two commuting Killing vectors.

\section{Spatially homogeneous models of Shapovalov spaces}

In this paper, we will consider spatially homogeneous wave models of space-time to describe primary gravitational perturbations. For Shapovalov's wave-like spacetimes to be spatially homogeneous, it is necessary. so that they admit three-parameter groups of motion with space-like orbits. As we showed earlier \cite{Osetrin2020RPJ410}, for Shapovalov spaces of type~II there are two types of spatially homogeneous models - subtypes B1 and B2.

\subsection{Spatially homogeneous wave-like model type~{II-B1} }

The metric for the space-time model of type {B1} has the form (see \cite{Osetrin2020RPJ410}):
\begin{equation}
ds^2=
\frac{1}{{x^3}^2}
\left(
2\,dx^0 dx^1+({x^0}-{\alpha})^{1-{\beta}} ({x^0}+{\alpha})^{1+{\beta}}\,{dx^2}^2+{dx^3}^2
\right),
\label{MetricSubB1}
\end{equation}
where $ x^0 $ is an isotropic (wave) variable, $\alpha$ and $\beta$ are constants ($\alpha\ne 0$, $\beta\ne\pm 1$).
$$
{g=\det g_{ij}=}-\frac{({x^0}-{\alpha})^{1-{\beta}} ({\alpha}+{x^0})^{{\beta}+1}}{{x^3}^8},
\qquad
x^0>|\alpha|.
$$
%
%
Killing vector fields of model type {B1} in a privileged coordinate system can be selected in the form:
\be
X_0=\partial_1,
\qquad
X_1=\partial_2,
\qquad
X_2=2\,x^1\partial_1+ x^2\partial_2+x^3\partial_3,
\ee
\begin{equation}
X_3= ({x^0}^2-\alpha^2)\partial_0-\frac{{x^3}^2}{2}\partial_1+ {\alpha} \beta x^2\partial_2+x^0 x^3\partial_3.
\end{equation}
Killing vectors $ X_1 $, $ X_2 $, $ X_3 $ define a subgroup of spatial homogeneity of the model.
Killing vector commutators of  model type {B1} have the form:
\be
[{X_0},{X_1}]=0,
\qquad
[{X_0},{X_2}]=2{X_0},
\qquad
[{X_0},{X_3}]=0,
\ee
\begin{equation}
[{X_1},{X_2}]={X_1},
\qquad
[{X_1},{X_3}]={\alpha}\beta{}{X_1},
\qquad
 [{X_2},{X_3}]=0.
\end{equation}
For $ \alpha\beta = 0 $, this space admits a third commuting Killing vector and degenerates into
a space with one non-ignored variable only in the privileged coordinate system.

The Riemann tensor $ R_{ijkl} $, the Ricci tensor $ R_{ij} $ and the scalar curvature~$R$ have the following nonzero components:
\begin{equation}  
R_{0313}=-\frac 1{{x^3}^4},
\quad
R_{0212}=R_{2323}=-\frac{(x^0-\alpha)^{1-\beta}(x^0+\alpha)^{1+\beta}}{{x^3}^4}, 
\end{equation} 
\begin{equation}  
R_{0202}=\frac{\alpha^2(1-\beta^2)(x^0-\alpha)^{-1-\beta}(x^0+\alpha)^{-1+\beta}}{{x^3}^2},
\end{equation} 
\begin{equation}  
R_{00}=\frac{\alpha^2(1-\beta^2)}{({x^0}^2-\alpha^2)^2},
\qquad
R_{01}=R_{33}=-\frac 3{{x^3}^2},
\end{equation} 
\begin{equation}  
R_{22}=-\frac{3\,(x^0-\alpha)^{1-\beta}(x^0+\alpha)^{1+\beta}
}{{x^3}^2},
\qquad
R=-12
.
\end{equation} 
If $ \alpha \ne 0 $ and $ \beta \ne \pm 1 $ the resulting solution cannot be conformally flat, since the two components of the Weyl tensor $C_{ijkl}$ are not equal to zero:
\begin{equation}
C_{0202} = \frac{{\alpha}^2 \left(1-{\beta}^2\right) ({x^0}-{\alpha})^{-{\beta}-1} ({x^0}+{\alpha})^{{\beta}-1}}{2 {x^3}^2},
\end{equation}
\begin{equation}
C_{0303} = \frac{-{\alpha}^2 \left(1-{\beta}^2\right)}{2 {x^3}^2 ({x^0}-\alpha)^2 ({x^0}+{\alpha})^2}
.
\end{equation}
If  $ {\alpha} = 0 $ or $\beta=0, \pm 1$, the metric of the model type~{B1} degenerates -
in a privileged coordinate system it depends on one variable only.

The model {B1} is of type {III} according to the Bianchi classification and has the type N according to the Petrov classification.

\subsection{Spatially homogeneous wave-like model  type~{B2}}

The space-time interval for a  type~{B2} model  can be written as (see \cite{Osetrin2020RPJ410}):
\begin{equation}
ds^2=\frac{1}{{x^3}^2}\,\left(
2\,dx^0dx^1+{x^0}^{\alpha}\,{dx^2}^2 +{dx^3}^2
\right),
\label{MetricSubB2}
\end{equation}
where $ x^0 $ is an isotropic (wave) variable, and ${\alpha}$ is a constant.
$$
{g=\det g_{ij}=}-{x^0}^{\alpha}/{x^3}^8,
\qquad
x^0>0
.
$$
Independent Killing vector fields in a privileged coordinate system can be selected in the form:
\be
X_0=\partial_1,
\qquad
X_1=\partial_2,
\qquad
X_2=2\,x^1 \partial_1+x^2\partial_2+x^3\partial_3,
\ee
\begin{equation}
X_3=x^0\partial_0+ \frac{1-\alpha}{2}\,x^2\partial_2+\frac{x^3}{2}\,\partial_3.
\end{equation}
Killing vectors $ X_1 $, $ X_2 $, $ X_3 $ define a subgroup of spatial homogeneity of the model.
Killing vector commutators of  model  type~{B2} have the form:
\be
[{X_0},{X_1}]=0,
\qquad
[{X_0},{X_2}]=2{X_0},
\qquad
[{X_0},{X_3}]=0,
\ee
\begin{equation} 
[{X_1},{X_2}]={X_1},
\qquad
[{X_1},{X_3}]=\frac{1-\alpha}{2}\,{X_1},
\qquad
[{X_2},{X_3}]=0.
\end{equation}
For $ {\alpha} =  1 $, this space admits an additional  commuting Killing vector and degenerates into a space with one non-ignored variable.

The Riemann tensor $ R_{ijkl} $, the Ricci tensor $ R_{ij} $ and the scalar curvature $ R $ have the following nonzero components:
\be
R_{0101}=-R_{0313}=\frac 1{{x^3}^4},\ \ \ R_{0212}=R_{2323}=-\frac{{x^0}^{\alpha}}{{x^3}^4},\ \ \ 
R_{0202}=\frac{\alpha(2-\alpha)}{4{x^3}^2{x^0}^{(2-\alpha)}}, 
\ee
\begin{equation} 
 R_{01}=R_{33}=-\frac 3{{x^3}^2},\ \ \ R_{22}=-\frac{3{x^0}^{\alpha}}{{x^3}^2},\ \ \ R_{00}=\frac{\alpha(2-\alpha)}{4{x^0}^2},\ \ \ R=-12
 .
 \end{equation} 
If $ \alpha \ne 0 $ or $ \alpha \ne 2 $  the resulting solution cannot be conformally flat, since the two components of the Weyl tensor $C_{ijkl}$ are not equal to zero:
\begin{equation} 
{\rm C}_{0202} =-\frac{{\alpha}\, ({\alpha}-2) {x^0}^{({\alpha}-2)}}{8 {x^3}^2},
\qquad
{\rm C}_{0303} =\frac{{\alpha}\, ({\alpha}-2)}{8 {x^0}^2 {x^3}^2}
 \end{equation} 
If $ \alpha = 0 $ or $ \alpha = 2 $, the Weyl tensor vanishes (conformally flat space), but the Ricci tensor, scalar curvature and the Riemann curvature tensor does not vanish. 
If  $ {\alpha} = 0, 1, 2 $, the metric of the model type~{B2} degenerates -
in a privileged coordinate system it depends on one variable only.

This spatially homogeneous space-time model is of type {III} according to the Bianchi classification and has type N according to the Petrov classification.

\section{Shapovalov spacetimes II-B1 type}

For the metric of the spatially homogeneous Shapovalov model of type II-B1, we write the explicit form of the field equations in vacuum
 ($T_{\alpha\beta}=0$):
\begin{eqnarray}
\mathbf{E_{00}}&=&
\frac{1}{\varkappa ^2}
\Biggl[
\frac{{\alpha}^2 \left({\beta}^2-1\right) (24 {\gamma}{}-1)}{\left({x^0}^2-{\alpha}^2\right)^2}+24
   \left({\varphi}_{,0}{}^2
   {\gamma}''{}+{\varphi}_{,00}{}
   {\gamma}'{}\right)
\Biggr]
\nonumber\\
&-&
\frac{1}{{({\alpha}^2-{x^0}^2)}^3 }
\biggl\{
8 {\alpha}^2 \left({\beta}^2-1\right) {x^3}^2 ({\alpha}^2-{x^0}^2) {\varphi}_{,3}{}^2 {\xi}''{}
\nonumber\\
&+&
8   
\biggl[
({\alpha} {\beta}-{x^0}) \left(\left({x^0}^2-{\alpha}^2\right)^2
   \left(-{\varphi}_{,0}{}\right)-{\alpha}^2 \left({\beta}^2-1\right) {x^3}^2
   {\varphi}_{,1}{}\right)  
\nonumber\\
&+&
{\alpha}^2 \left({\beta}^2-1\right) {x^3}^2 ({\alpha}^2-{x^0}^2)
   {\varphi}_{,33}{}
\nonumber\\
&+&
{\alpha}^2 \left({\beta}^2-1\right) {x^3} ({\alpha}^2-{x^0}^2)
   {\varphi}_{,3}{}+\left({\alpha}^2-{x^0}^2\right)^3 {\varphi}_{,00}{}
\biggr]
   {\xi}'{}
\nonumber\\
&+&
8 {\alpha}^2 \left({\beta}^2-1\right) ({\alpha}^2-{x^0}^2)
   {\xi}{}+\left({\alpha}^2-{x^0}^2\right)^3 {\varphi}_{,0}{}^2 \left(8
   {\xi}''{}+{\omega}\right)
\biggr\}
=0
,
\label{B1EqField00}
\end{eqnarray}
\begin{eqnarray}
\mathbf{E_{01}}
&=&
-{x^3} 
\biggl[
-{x^3} 
\Bigl\{
\left({\alpha}^2-{x^0}^2\right) ({\alpha}+{x^0})^{{\beta}} {\varphi}_{,3}{}^2 \left(\varkappa ^2 \left(16 {\xi}''+{\omega}\right)
-48 {\gamma}''{}\right)
\nonumber\\
&-&
16 \left({\alpha}^2-{x^0}^2\right) ({\alpha}+{x^0})^{{\beta}} {\varphi}_{,1}{\varphi}_{,0}\left(3 {\gamma}''{}-\varkappa ^2 {\xi}''\right)
\nonumber\\
&+&
({x^0}-{\alpha})^{{\beta}} {\varphi}_{,2}{}^2 
\left(-\left(\varkappa ^2 \left(16 {\xi}''{}+{\omega}\right)-48 {\gamma}''
\right)\right)
\Bigr\}
\nonumber\\
&-&
48({\alpha}+{x^0})^{{\beta}}
\Bigl[
 \left({\alpha}^2-{x^0}^2\right) 
\left( {\varphi}_{,3}{}- {x^3}{\varphi}_{,33} \right)
\nonumber\\
&-&
{x^3} \left(
\left(\left({\alpha}^2-{x^0}^2\right) {\varphi}_{,01}{}
+({\alpha} {\beta}-{x^0}) {\varphi}_{,1}{}\right)
-
{\varphi}_{,22}{}\right)
\Bigr]
{\gamma}'{}
\nonumber\\
&+&
16 \varkappa ^2 
\Bigl\{
-{x^3}
\biggl( 
- ({x^0}-{\alpha})^{{\beta}} {\varphi}_{,22}
+
\left({\alpha}^2-{x^0}^2\right) ({\alpha}+{x^0})^{{\beta}} {\varphi}_{,33}{}
\nonumber\\
&+&
({\alpha}+{x^0})^{{\beta}} 
\left(\left({\alpha}^2-{x^0}^2\right) {\varphi}_{,01}{}+2 ({\alpha} {\beta}-{x^0}) {\varphi}_{,1}{}\right)
\biggr)
\nonumber\\
&+&
2 \left({\alpha}^2-{x^0}^2\right) ({\alpha}+{x^0})^{{\beta}} {\varphi}_{,3}{}
\Bigr\} {\xi}'{}
\,
\biggr]
\nonumber\\
&+&
2 \varkappa ^2 \left({\alpha}^2-{x^0}^2\right) ({\alpha}+{x^0})^{{\beta}}(60 {\xi}{}+{\upsilon} )
=0
,
\label{B1EqField01}
\end{eqnarray}
\begin{eqnarray}
\mathbf{E_{02}}
&=&
24 \left(\left({\alpha}^2-{x^0}^2\right) {\varphi}_{,02}{}+({x^0}-{\alpha} {\beta}) {\varphi}_{,2}{}\right) {\gamma}'{}
\nonumber\\
&-&
8 \varkappa ^2 \left(\left({\alpha}^2-{x^0}^2\right) 
{\varphi}_{,02}{}+({x^0}-{\alpha} {\beta}) {\varphi}_{,2}{}\right) {\xi}'{}
\nonumber\\
&-&
\left({\alpha}^2-{x^0}^2\right) {\varphi}_{,2}{} {\varphi}_{,0}{} 
\left(\varkappa ^2 \left(8 {\xi}''
+{\omega}\right)-24{\gamma}''{}\right)
=0
,
\label{B1EqField02}
\end{eqnarray}
\begin{eqnarray}
\mathbf{E_{03}}
&=&
\frac{1}{{x^3}}
\,
\biggl\{
\,
\frac{24 }{\varkappa ^2}
\left({x^3} {\varphi}_{,3}{} {\varphi}_{,0}{} 
{\gamma}''{}+\left({\varphi}_{,0}{}+{x^3} {\varphi}_{,03}{}\right) 
{\gamma}'{}\right)
\nonumber\\
&-&
\frac{1}{\left({x^0}^2-{\alpha}^2\right)^2}
\biggl[
{x^3} {\varphi}_{,3}{} 
\Bigl\{
8 \left(\left({x^0}^2-{\alpha}^2\right)^2 {\varphi}_{,0}{}-{\alpha}^2 \left({\beta}^2-1\right) {x^3}^2 {\varphi}_{,1}{}\right){\xi}''
\nonumber\\
&+&
{\omega} \left({x^0}^2-{\alpha}^2\right)^2 {\varphi}_{,0}{}
\Bigr\}
+8 
\Bigl(
\left({x^0}^2-{\alpha}^2\right)^2 \left({\varphi}_{,0}{}+{x^3} {\varphi}_{,03}{}\right)
\nonumber\\
&-&
{\alpha}^2 \left({\beta}^2-1\right) {x^3}^2
\left(
{x^3} {\varphi}_{,13}
-{\varphi}_{,1}
\right)
\Bigr) {\xi}'{}
\biggr]
\biggr\}
=0
,
\label{B1EqField03}
\end{eqnarray}
\begin{eqnarray}
\mathbf{E_{11}} & = &
8 {\varphi}_{,11}{} \left(3 {\gamma}'{}-\varkappa ^2 {\xi}'{}\right)
-{\varphi}_{,1}{}^2 
\left[
\varkappa ^2 \left(8 {\xi}''{}+{\omega}\right)-24 {\gamma}''{}
\right]
=0
,
\label{B1EqField11}
\end{eqnarray}
\begin{eqnarray}
\mathbf{E_{12}} &= &
8 {\varphi}_{,12}{} \left(3 {\gamma}'{}-\varkappa ^2 {\xi}'{}\right)
-{\varphi}_{,1}{} {\varphi}_{,2}{}
\left[
\varkappa ^2 \left(8 {\xi}''{}+{\omega}\right)-24 {\gamma}''{}
\right]
=0
,
\label{B1EqField12}
\end{eqnarray}
\begin{eqnarray}
\mathbf{E_{13}}
&=&
24 \left({\varphi}_{,1}{}+{x^3} {\varphi}_{,13}{}\right) {\gamma}'{}
\nonumber
\\
&-&
{x^3} 
\left[
{\varphi}_{,3}{} {\varphi}_{,1}{} \
\left(\varkappa ^2 \left(8 {\xi}''{}+{\omega}\right)-24 {\gamma}''{}\right)+8 \varkappa ^2 \
{\varphi}_{,13}{} {\xi}'{}
\right]
=0
,
\label{B1EqField13}
\end{eqnarray}
\begin{eqnarray}
\mathbf{E_{22}} &=&
{x^3} 
\biggl\{
\,
{x^3} 
\,
\Bigr[
\,
\left({x^0}^2-{\alpha}^2\right) ({\alpha}+{x^0})^{{\beta}} 
{\varphi}_{,3}{}^2 \left(\varkappa ^2 \left(16 {\xi}''{}+{\omega}\right)
-48 {\gamma}''{}\right)
\nonumber
\\
&+&
2 \left({x^0}^2-{\alpha}^2\right) ({\alpha}+{x^0})^{{\beta}} {\varphi}_{,1}{} {\varphi}_{,0}{} 
\left(\varkappa ^2 \left(16 {\xi}''{}+{\omega}\right)-48 {\gamma}''{}\right)
\nonumber
\\
&-&
{\omega} \varkappa ^2 ({x^0}-{\alpha})^{{\beta}} {\varphi}_{,2}{}^2
\,
\Bigr]
\nonumber
\\
&-&
48 \left({\alpha}^2-{x^0}^2\right) ({\alpha}+{x^0})^{{\beta}} 
\left({\varphi}_{,3}{}-{x^3} \left({\varphi}_{,33}{}
+2 {\varphi}_{,01}{}
\right)\right) {\gamma}'{}
\nonumber
\\
&+&
16 \varkappa ^2 ({\alpha}+{x^0})^{{\beta}} 
\Bigl[
{x^3} 
\Bigl(
\left({x^0}^2-{\alpha}^2\right) {\varphi}_{,33}{}
+2 \left({x^0}^2-{\alpha}^2\right) {\varphi}_{,01}{}
\nonumber
\\
&+&
({\alpha} {\beta}-{x^0}) {\varphi}_{,1}{}
\Bigl)
+2 \left({\alpha}^2-{x^0}^2\right) {\varphi}_{,3}{}
\Bigr] \, {\xi}'{}
\,
\biggr\}
\nonumber
\\
&-&
2 \varkappa ^2 \left({\alpha}^2-{x^0}^2\right) ({\alpha}+{x^0})^{{\beta}} 
\left(
{\upsilon}
+60{\xi}
\right)
=0
,
\label{B1EqField22}
\end{eqnarray}
\begin{eqnarray}
\mathbf{E_{23}}&=&
{x^3} {\varphi}_{,3}{} {\varphi}_{,2}{} 
\left[
24 {\gamma}''{}
-\varkappa ^2 \left(8 {\xi}''{}+{\omega}\right)
\right]
\nonumber
\\
&+&
24 \left({\varphi}_{,2}{}+{x^3} {\varphi}_{,23}{}\right) {\gamma}'{}
+8 \varkappa ^2 \left({\varphi}_{,2}{}-{x^3} {\varphi}_{,23}{}\right) {\xi}'{}
=0
,
\label{B1EqField23}
\end{eqnarray}
\begin{eqnarray}
\mathbf{E_{33}}
&=&
2 \varkappa ^2 ({\alpha}+{x^0})^{{\beta}} \left({x^0}^2-{\alpha}^2\right)^2  
\left(
{\upsilon}{}+60 {\xi}
\right)
\nonumber
\\
&+&
{x^3} 
\,
\biggr\{
{x^3} 
\Biggl(
-2 
\biggl( 
8 \varkappa ^2 
\Bigl(
({x^0}-{\alpha})^{{\beta}} \left({\alpha}^2-{x^0}^2\right) 
{\varphi}_{,22}
\nonumber
\\
&+&
({\alpha}+{x^0})^{{\beta}} 
\Bigl[
{\alpha}^2 \left({\beta}^2-1\right) {x^3}^2 {\varphi}_{,11}{}
-\left({\alpha}^2-{x^0}^2\right) ({\alpha} {\beta}-{x^0}) {\varphi}_{,1}{}
\nonumber
\\
&-&
2 \left({x^0}^2-{\alpha}^2\right)^2 {\varphi}_{,01}{}
\Bigr]
\left.
\left.
\Bigr) 
\,
{\xi}'{}
\right.
\right.
\nonumber
\\
&-&
({\alpha}+{x^0})^{{\beta}} {\varphi}_{,1}{} 
\left(
8 \varkappa ^2 
\left[ 2 
\left({x^0}^2-{\alpha}^2\right)^2 {\varphi}_{,0}{}
-{\alpha}^2 
\left({\beta}^2-1\right) {x^3}^2 {\varphi}_{,1}{}
\right] {\xi}''{}
\right.
\nonumber
\\
&+&
\left.
\left.
\left({x^0}^2-{\alpha}^2\right)^2 {\varphi}_{,0}{} \left({\omega} \varkappa ^2
-48 {\gamma}''{}
\right)\right)\right)
\nonumber
\\
&-&
\left({\alpha}^2-{x^0}^2\right) ({x^0}-{\alpha})^{{\beta}} {\varphi}_{,2}{}^2 \left(\varkappa ^2 \left(16 {\xi}''{}+{\omega}\right)
-48 {\gamma}''{}\right)
\nonumber
\\
&-&
\left.
{\omega} \varkappa ^2 \left({x^0}^2-{\alpha}^2\right)^2 ({\alpha}+{x^0})^{{\beta}} {\varphi}_{,3}{}^2\right)
\nonumber
\\
&+&
48 \left({\alpha}^2-{x^0}^2\right) 
\biggl[
3 \left({\alpha}^2-{x^0}^2\right) ({\alpha}+{x^0})^{{\beta}} {\varphi}_{,3}{}
\nonumber
\\
&+&
{x^3}
\Bigl(
({\alpha}+{x^0})^{{\beta}} \left(2 \left({x^0}^2-{\alpha}^2\right) {\varphi}_{,01}{}
+({x^0}-{\alpha} {\beta}) {\varphi}_{,1}{}\right)
\nonumber
\\
&+&
({x^0}-{\alpha})^{{\beta}} {\varphi}_{,22}
\,
\Bigr)
\,
\biggr] 
\,
{\gamma}'
\,
\biggr\}
=0
.
\label{B1EqField33}
\end{eqnarray}
The equation (\ref{EqScalar}) for the scalar field  takes the form
$$
\frac{{\omega}}{{\alpha}^2-{x^0}^2}
\biggl\{
{x^3} ({\alpha}+{x^0})^{-{\beta}} 
\Bigl(
{x^3}
\Bigl[
\left({\alpha}^2-{x^0}^2\right) ({\alpha}+{x^0})^{{\beta}} {\varphi}_{,33}{}
$$
$$
\mbox{}
+({\alpha}+{x^0})^{{\beta}} \left(2 \left({\alpha}^2-{x^0}^2\right) {\varphi}_{,01}{}
+({\alpha} {\beta}-{x^0}) {\varphi}_{,1}{}\right)
$$
\be
\mbox{}
-({x^0}-{\alpha})^{{\beta}} {\varphi}_{,22}
\,
\Bigr]
-2 \left({\alpha}^2-{x^0}^2\right) ({\alpha}+{x^0})^{{\beta}} {\varphi}_{,3}
\Bigr)
\biggr\}
=
{\upsilon}'{}+84 {\xi}'{}
-\frac{72 {\gamma}'{}}{\varkappa ^2}
.
\label{B1EqScalar}
\ee
The "privileged" coordinate system used allows complete separation of variables in the Hamilton-Jacobi equation for test particles.
We will seek the solution of the field equations in the class of scalar fields having the ''split'' form
\be
\varphi(x^0,x^1,x^2,x^3)=\varphi_0(x^0)\varphi_1(x^1)\varphi_2(x^2)\varphi_3(x^3)
.
\ee

Accordingly, there are cases when the scalar field in the used ''privileged'' coordinate system depends on the ignored variables $ x^1 $ and $ x^2 $ (on which the metric does not depend) and the case when the scalar field depends only on the non-ignored variables $ x^0 $ and $ x^3 $. Moreover, the coordinates $ x^0 $ and $ x^1 $ are isotropic (along them the space-time interval is equal to zero).

\subsection{Spacetimes II-B1 type. Scalar field depends on ignored variables.}


The structure of the field equations of the quadratic theory of gravity makes it possible to isolate the dependence of the scalar field $ \varphi $ on the ignored variables $ x^1 $ and $ x^2 $, and the variable $ x^1 $, which is isotropic, is isolated, and it is convenient to start the search for solutions by studying this case, when the scalar field depends on the variable $ x^1 $. We will show below that this case does not admit solutions, i.e. the scalar field can only depend on the non-ignored wave variable $ x^0 $. The case when the scalar field depends on the second, ignored, but non-isotropic variable $ x^2 $ is also considered below and it leads to an exact solution for the considered model.

\subsubsection{Spacetimes II-B1 type, case ${\varphi}_{,1}\ne 0$.}
The space-time metric for spatially homogeneous Shapovalov wave models of type II-B1 has the form
(\ref{MetricSubB1}):
$$
ds^2=
\frac{1}{{x^3}^2}
\left(
2\,dx^0 dx^1+({x^0}-{\alpha})^{1-{\beta}} ({x^0}+{\alpha})^{1+{\beta}}\,{dx^2}^2+{dx^3}^2
\right),
$$
where $ x^0 $ is an isotropic (wave) variable, $\alpha$ and $\beta$ are constants ($\alpha\ne 0$, $\beta\ne\pm 1$).


Consider first the case where $ \varphi $ depends on an ignored isotropic variable
$ x^1 $, i.e. $ \varphi_{, 1} \ne 0 $.
Then from the equation (\ref{B1EqField11}) we get two consequences:
$$
{\varphi_1}'=\rho\, {\varphi_1}^{\sigma}
,\qquad
\rho,{\sigma}
-\mbox{const}.
$$
\be
\gamma''=
\frac{
{\omega} \varkappa ^2 \varphi -24 {\sigma} {\gamma}'(\varphi )+8 {\sigma} \varkappa ^2 {\xi}'(\varphi )+8 \varkappa ^2 \varphi  {\xi}''(\varphi )
   }{24 \varphi }
   ,
 \label{B1gamma2}
\ee
Then, in the considered case ($ {\varphi_1}' \ne 0 $), the equation (\ref{B1EqField12}) implies
\be
({\sigma}-1)\,{\varphi_2}'=0
,
\ee
moreover, for $ {\sigma} = 1 $, additionally from the equation (\ref{B1EqField02}) it follows $ {\varphi_2}'= 0 $, and from the equation (\ref{B1EqField13}) we have $ \gamma = \mbox{const } $.
Thus, for $ {\varphi_1}'\ne 0 $, it always follows from the field equations:
\be
{\varphi_2}'=0
.
\ee
Then from the equation (\ref{B1EqField13}) we get:
\be
\frac{\gamma'}{\gamma'-\varkappa^2\xi'/3}=\mbox{const}=\tau,
\qquad
({\sigma}-1)x^3{\varphi_3}'=\tau\varphi_3
.
\label{B1phi3}
\ee
Consider the case $ {\sigma} = 1 $, then $ \tau = 0 $ and $ \gamma'= 0 $, and the equation (\ref{B1gamma2}) implies
\be
{\omega}  \varphi/8 +  {\xi}'(\varphi )+  \varphi\,  {\xi}''(\varphi )=0
\quad \to \quad
\xi(\varphi )=
c_1 \ln (\varphi )-\frac{\omega }{32}\, \varphi^2+c_2
.
\ee
Then the equation (\ref{B1EqField03}) implies $ \alpha (1- \beta ^ 2) = {\varphi_0}'= 0 $, i.e. this is the degenerate case of a conformally flat space.

We will now assume that ${\sigma}\ne 1$.
Then from (\ref{B1phi3}) for $ {\sigma} \ne 1 $ we obtain the following corollaries:
\be
{\varphi_3}'=\frac{\tau\varphi_3}{({\sigma}-1)x^3},
\qquad
{\varphi_3}(x^3)=c_3 {x^3}^{\tau/({\sigma}-1)}
,\qquad
\xi'=\frac{3(\tau-1)}{\varkappa^2\tau}\gamma'
.
\ee
Substitution of these conditions into the remaining equations leads to their inconsistency and the absence of solutions. Thus, we have shown that a scalar field cannot depend on an ignored isotropic variable $x^1$.

\subsubsection{Solution for spacetimes II-B1 type for ${\varphi_1}'=0$, ${\varphi_2}'\ne 0$.}

Let us now consider the case when the scalar field does not depend on the ignored isotropic variable $ x^1 $, but depends on the second ignored (non-isotropic) variable $ x^2 $:
\be
\varphi=\varphi (x^0,x^2,x^3)=\varphi_0(x^0)\varphi_2(x^2)\varphi_3(x^3),
\qquad
\varphi_2'\ne 0
.
\ee
Then the compatibility condition for the equations (\ref{B1EqField03}) and (\ref{B1EqField23}) leads to the requirement
$\xi=\mbox{const}$
and additionally
\be
\varphi_3(x^3)=c_1\,{x^3}^{\sigma},
\qquad
{c_1}, {\sigma} - \mbox{const}
.
\ee
For the function $ \gamma (\varphi) $ we obtain from (\ref{B1EqField03}) and (\ref{B1EqField23}) the differential equation
\be
\gamma''(\varphi)=
\frac{\varkappa ^2 \omega }{24}-\frac{1+1/{{\sigma} } }{\varphi }
\, \gamma' (\varphi )
.
\label{B1gammaEq}
\ee
The equation (\ref{B1gammaEq}) has two types of solutions depending on the value of the constant~${\sigma}$:
\be
{\sigma}\ne 0,-1/2
\quad
\to
\quad
\gamma (\varphi)=
\frac{
\omega{\sigma} \varkappa^2 
}{48 (2 {\sigma} +1)}\, \varphi^{2}
+c_2 \varphi^{-1/{\sigma} }
+c_3
.
\label{B1gammaSol1}
\ee
\be
{\sigma}=-1/2
\quad
\to
\quad
\gamma (\varphi)= 
\frac{\varkappa ^2 \omega}{48}   \varphi ^2 \ln (\varphi )
+c_4 \varphi ^2+c_5
,
\label{B1gammaSol2}
\ee

Let us first consider the second type of solutions
-- the case (\ref{B1gammaSol2}), when \mbox{${\sigma}=-1/2$}.
Then from the equation~(\ref{B1EqField02}) we obtain
\be
\varphi_0(x^0)=
({x^0}-{\alpha})^{\frac{1-{\beta}}{4}}
   ({x^0}+{\alpha})^{\frac{1+{\beta}}{4}}
   ,
\label{B1gammaSol2phi0}
\ee
Substituting the function $ \varphi_0 (x^0) $ from (\ref{B1gammaSol2phi0}) into the rest of the equations, we obtain from (\ref{B1EqField00}) the condition
\be
\alpha (\beta^2-1)=0
,
\ee
those this option leads to degeneration (the space becomes conformally flat) and the absence of solutions.

Consider now the case (\ref{B1gammaSol1}), when $ {\sigma} \ne 0, -1/2 $.
Then integrating the equation~(\ref{B1EqField02}), we obtain
\be
\varphi_0(x^0)=
c_4 \left[ \frac{(x^0-\alpha)^{\beta-1}}{(x^0+\alpha)^{\beta+1}} \right]^{{\sigma}/2}
.
\ee
The equation (\ref{B1EqField00}) with $ {\sigma} \ne -1 / 2 $ becomes an identity only if
\be
c_2=0
,\qquad
c_3=(1+8\varkappa^2\xi)/24
,\qquad
{\sigma}=1/2
.
\ee
The field equations (\ref{B1EqField01}), (\ref{B1EqField22}) and (\ref{B1EqField33}) define $ \varphi_2 (x^2) $ and the form of the scalar potential $ V (\varphi) $, thereby giving complete solution of the problem under consideration.

Thus, we obtain an exact solution of the field equations of the quadratic theory of gravity with a scalar field for a spatially homogeneous wave-like model of Shapovalov space of type II-B1:
$$
ds^2=
\frac{1}{{x^3}^2}
\left(
2\,dx^0 dx^1+({x^0}-{\alpha})^{1-{\beta}} ({x^0}+{\alpha})^{1+{\beta}}\,{dx^2}^2+{dx^3}^2
\right)
,
$$
\be
\xi(\varphi ) = {\xi}=\mbox{const}
,\qquad
\gamma(\varphi ) = \frac{{\omega} \varkappa ^2}{192}\, {\varphi}^2+\frac{1}{24} \left(1+8 {\xi} \varkappa ^2\right)
,
\ee
\be
V(\varphi ) =
\frac{\omega\rho^2}{8{k}^4}\, {\varphi}^6 
-\frac{{\omega}}{4} {\varphi}^2
-60 {\xi}-\frac{3}{\varkappa ^2}
,
\ee
\be
\varphi=
\varphi(x^0,x^2,x^3) = 
{k}
\,
\frac{({x^0}-{\alpha})^{({\beta}-1)/4}}{({x^0}+{\alpha})^{({\beta}+1)/4} }
\sqrt{
\frac{
{x^3} 
}{ { {\rho} {x^2}+{\delta} }}
}
,
\ee
where $ x^0 $ is an isotropic wave variable, $ \alpha $ and $ \beta $ are constant parameters of the space-time model,
$ {k} $, $ {\rho} $, $ {\delta} $ are constants of integration.

Note that in the obtained solution, the term in the Lagrangian associated with the Gauss-Bonnet invariant does not affect the dynamics of the model -- it contributes only to the cosmological constant. The presence of a dependence of the scalar field on the ignored variable $ x^2 $ leads to the appearance in the scalar potential $ V (\varphi) $ of terms including the
$ {\varphi}^6 $. The constant $ \Lambda = 60 {\xi} + {3} / {\varkappa^2} $ plays the role of the cosmological constant in the obtained solution. The solution can describe the primary gravitational-wave disturbances in a spatially homogeneous non-isotropic Universe at the early stages of its development up to the stage of "isotropization".

\subsection{Spacetimes II-B1 type. Scalar field depends on non-ignored variables only.}

In this subsection, we will consider the case when the scalar field $ \varphi $ in the ''privileged'' coordinate system we use depends only on the non-ignored variables $ x^0 $ and $ x^3 $ (on which the space-time metric depends).
\be
\phi=\phi(x^0,x^3)=\phi_0(x^0)\,\phi_3(x^3)
.
\ee
Moreover, the variable $ x^0 $ is a wave isotropic variable (along it the space-time interval is equal to zero).
The obtained solutions in the framework of the quadratic theory of gravity, taking into account the fact that the considered model is spatially homogeneous, give us models of wave gravitational perturbations at the early stages of the development of the Universe -- these are examples of exact solutions for primary gravitational waves.

From the field equation (\ref{B1EqField33}) it follows that
\be
\phi_3(x^3)={x^3}^{\sigma}
,\qquad
{\sigma} - \mbox{const}
,
\ee
\be
\frac{72{\sigma} }{\varkappa ^2} \phi {\gamma}'(\phi)=
\frac{{\omega}{\sigma}^2}{2} \phi^2
-{{\upsilon}(\phi)}-{60 {\xi}(\phi)}
.
\label{B1gammaDerivativ}
\ee
Substituting the form of the derivative $ {\gamma}' (\phi) $ from (\ref{B1gammaDerivativ}) into the field equations, we obtain from the compatibility conditions for the equations (\ref{B1EqField01}), (\ref{B1EqField03}) and~(\ref{B1EqField22})
\be
\xi(\phi)=\xi=\mbox{const}
.
\ee
Then from the equations (\ref{B1EqField01}), (\ref{B1EqField03}), (\ref{B1EqField22}) there follow two solutions depending on the value of the constant $ \sigma $. For $ \sigma = -1 / 2 $ - one solution, for $ \sigma \ne -1/2 $ - another solution.

Consider first the special case $ \phi = \phi (x^0, x^3) $ for $ \sigma = -1 / 2 $. Then from the equation (\ref{B1EqField33}) we obtain for the scalar field the following form
\be
\phi(x^0,x^3) = 
\frac{
{\phi_0}(x^0)
}{
\sqrt{x^3}
}
.
\ee
The compatibility conditions for the equations (\ref{B1EqField01}), (\ref{B1EqField03}), (\ref{B1EqField22}) and the integration of these equations yield
\be
\gamma(\phi)=
\frac{1}{72} c_1 \kappa ^2 \phi^2+\frac{\kappa ^2}{576}  \left[ 12 {\omega} \phi^2 \ln (\phi) -7 {\omega} \phi^2-480
   {\xi}\right]
   +c_2
   ,
\ee
\be
V(\phi)=
c_1 \phi^2+\frac{3}{2} {\omega} \phi^2 \ln (\phi)-60 {\xi}
-\frac{3}{\kappa ^2}
.
\ee
Substituting this kind of scalar field functions into the remaining equation (\ref{B1EqField22}), we obtain as a consequence $ \alpha (1- \beta^2) = 0 $. Thus, in the considered variant of solving the problem for $ \sigma = -1 / 2 $, the space-time becomes ''degenerate'' -- conformally flat.

For completeness, we present the resulting conformally flat solution of the field equations:
$$
ds^2=
\frac{1}{{x^3}^2}
\left(
2\,dx^0 dx^1+({x^0}-{\alpha})^{1-{\beta}} ({x^0}+{\alpha})^{1+{\beta}}\,{dx^2}^2+{dx^3}^2
\right)
,\quad
\alpha(1-\beta^2)=0,
$$
$$
\phi(x^0,x^3)=c_1\sqrt{\frac{c_2 x^0-c_3}{x^3}},
\qquad
c_1,c_2, c_3,c_4 - \mbox{const}
,
$$
$$
\xi(\phi ) = {\xi}=\mbox{const}
,
$$
$$
\gamma(\phi ) =
\frac{1}{576} \left[
\kappa ^2 {\phi}{}^2 (12 {\omega} \ln ({\phi}{})+8 {c_4}-7  {\omega})+24 \left(8 {\xi} \kappa ^2+1\right)
\right],
$$
$$
V(\phi ) = {\phi}{}^2 \left(\frac{3}{2} {\omega} \ln ({\phi}{})+{c_4}\right)-60 {\xi}-\frac{3}{\kappa ^2}
.
$$

Let us now consider the main solution for $ \sigma \ne -1/2 $. Then the equations (\ref{B1EqField01}), (\ref{B1EqField03}) and (\ref{B1EqField22}) give
\be
\upsilon'(\phi)=
\left[ -120 \xi
+\omega \sigma^2 ( 2 \sigma-5) \phi^2
   - 2 \upsilon(\phi) \right]
/ (2 \sigma \phi)
 .
\label{B1upsilonDerivativ}
\ee
Integrating the equations (\ref{B1gammaDerivativ}) and (\ref{B1upsilonDerivativ}) we get
\be
\gamma(\phi ) = \frac{{c_1} \kappa^2}{72}  {\phi}{}^{-1/{\sigma}}+\frac{\kappa ^2 \left({\omega} {\sigma} {\phi}{}^2-40 (2 {\sigma}+1) \
{\xi}\right)}{48(2 {\sigma}+1)}+{c_2}
,
\qquad
\sigma\ne -1/2,
\ee
\be
V(\phi ) 
=
{c_1} {\phi}{}^{-1/{\sigma}}
+\frac{{\omega} {\sigma}^2 (2 {\sigma}-5) {\phi}{}^2}{4 {\sigma}+2}
-60 {\xi}
-\frac{3}{\kappa ^2}
,\qquad
c_1,c_2 - \mbox{const}
.
\ee
The remaining equation (\ref{B1EqField00}) expanded in powers of $ \phi $ gives the following conditions:
\be
\frac{{\alpha}^2 \left({\beta}^2-1\right) \left(
24 {c_2}-28 {\xi}   \kappa ^2-1
\right)
}{\kappa ^2 \left({x^0}^2-{\alpha}^2\right)^2}
=0,
 \label{B1Eq00A}
\ee
\be
{c_1} {\phi}{}^{-1/{\sigma}}
\,
\frac{
{\alpha}^2 \left({\beta}^2-1\right) {\sigma}^2{\phi_0}{}^2
+\left({
({\sigma}+1) {\phi_0}'{}^2
-{\sigma} {\phi_0}{} {\phi_0}''{}
}\right)
\left({x^0}^2-{\alpha}^2\right)^2
}{3   {\sigma}^2 \left({x^0}^2-{\alpha}^2\right)^2 {\phi_0}{}^2}
=0   ,
 \label{B1Eq00B}
\ee
\be
{\omega} 
\phi^2
\,
\frac{
{\alpha}^2 \left({\beta}^2-1\right)   {\sigma} {\phi_0}{}^2/2
+\left({
 {\sigma} {\phi_0}{}   {\phi_0}''{}
 - ({\sigma}+1)  {\phi_0}'{}^2
}\right)
\left({x^0}^2-{\alpha}^2\right)^2 
}{2 (2   {\sigma}+1) \left({x^0}^2-{\alpha}^2\right)^2 {\phi_0}{}^2}
 =0.
 \label{B1Eq00C}
\ee
From the equations (\ref{B1Eq00A}), (\ref{B1Eq00B}) and (\ref{B1Eq00C}) we obtain the following consequences:
\be
c_1=0
,\qquad
c_2=(1+28 {\xi}   \kappa ^2)/24
,
\ee
\begin{equation}
{\phi_0}{}   {\phi_0}''{}
- 
\frac{
({\sigma}+1)
}{{\sigma}}
\,
{\phi_0}'{}^2
+
\frac{
{\alpha}^2 \left({\beta}^2-1\right)
}{2\left({x^0}^2-{\alpha}^2\right)^2}
\,
{\phi_0}{}^2
=0
.
\label{B1phi0}
\end{equation}
The equation (\ref{B1phi0}) reduces to the Riccati equation 
\be
Z'(x^0)-\frac{1}{{\sigma}}Z^2(x^0)=\frac{ {\alpha}^2 \left(1-{\beta}^2\right) }{2 \left({x^0}^2-{\alpha}^2\right)^2},
\qquad
Z(x^0)=\frac{d \ln {\phi_0} }{d x^0}
 .
 \label{B1ZDerivativ}
\ee
Thus, we obtain an exact solution of the field equations of the quadratic theory of gravity
with scalar field
for a spatially homogeneous wave-like Shapovalov spacetime of type II-B1 of the form:
$$
ds^2=
\frac{1}{{x^3}^2}
\left(
2\,dx^0 dx^1+({x^0}-{\alpha})^{1-{\beta}} ({x^0}+{\alpha})^{1+{\beta}}\,{dx^2}^2+{dx^3}^2
\right),
$$
\be
\phi = 
\phi(x^0,x^3)={x^3}^{\sigma}\exp \int{Z(x^0)}\,dx^0
,
\qquad
{\sigma}\ne -1/2
,
\ee
where the function $ Z (x^0) $ is defined by the Riccati equation (\ref{B1ZDerivativ}), and 
scalar field functions from the Lagrangian~(\ref{Lagrangian}) take the form:
\be
\xi(\phi ) = {\xi}=\mbox{const},
\ee
\be
\gamma(\phi ) = 
\frac{\kappa ^2 {\omega} {\sigma}  }{48(2 {\sigma}+1)}\,
{{\phi}({x^0},{x^3})}^2
%
+(1+8  \kappa^2 {\xi})/24
,
\ee
\be
V(\phi ) = \frac{{\omega} {\sigma}^2 (2 {\sigma}-5) }{2(2 {\sigma}+1)}\,{{\phi}({x^0},{x^3})}^2
-60 {\xi}-{3}/{\kappa ^2}.
\ee
The constant $ \Lambda = 3 (20 {\xi} + {1} / {\kappa^2}) $ plays the role of the cosmological constant.
The term associated with the Gauss-Bonnet invariant contributes only to the comological constant.

\section{Shapovalov spacetimes II-B2 type}

Let us now consider the subtype II-B2 of spatially homogeneous Shapovalov models.
Let us write down the explicit form of the field equations for the type II-B2 metric:
\be
ds^2= \frac{1}{{x^3}^2}\left(
2\,dx^0dx^2
+{x^0}^{-{\alpha}}\,{dx^2}^2+{dx^3}^2
\right)
.
\label{MetricB2}
\ee
For the metric (\ref{MetricB2}), the field equations (\ref{EqsField}) take the form ($T_{\alpha\beta}=0$):
$$
\mathbf{E_{00}}=
\frac{{x^0}^3}{4\varkappa^2}
\left(
\frac{{\alpha} ({\alpha}+2) (1-24 {\gamma}{})}{ {x^0}^2}
-6 \left({\varphi}_{,0}{}^2 {\gamma}''{}+{\varphi}_{,00}{} {\gamma}'{}\right)
\right)
+2 {\alpha} ({\alpha}+2) {x^0} 
\,{\xi}{}
$$
$$
\mbox{}
+
 {\xi}'{}\,
\Bigl[
4 {\alpha} {x^0}^2
\left(
2 {x^0}{\varphi}_{,00}{}
+ {\varphi}_{,0}{}
\right)
+
{\alpha}^2 {x^3}^2
\left({\alpha} +2 \right) {\varphi}_{,1}{}
+2 {\alpha} {x^0} {x^3}({\alpha}+2)  
\left(
{x^3} {\varphi}_{,33}{}
+{\varphi}_{,3}{}
\right)
\Bigr]
$$
\be
\mbox{}
+
{x^0} 
\left[
2 \left({\alpha} ({\alpha}+2) {x^3}^2 {\varphi}_{,3}{}^2+4 {x^0}^2 
{\varphi}_{,0}{}^2\right) {\xi}''{}
+{\omega} {x^0}^2 {\varphi}_{,0}{}^2
\right]
=0
,
\label{B2EqField00}
\ee
$$
\mathbf{E_{01}}=
{x^0} {x^3} 
\,
\Bigl[
\,
{x^0}^{\alpha} {\varphi}_{,2}{}^2 
\left(\varkappa ^2 \left(16 {\xi}''{}+{\omega}\right)-48 {\gamma}''{}\right)
+{\varphi}_{,3}{}^2 \left(\varkappa ^2 \left(16 {\xi}''{}+{\omega}\right)-48 {\gamma}''{}\right)
$$
$$
\mbox{}
+16 {\varphi}_{,1}{} {\varphi}_{,0}{} \left(\varkappa ^2 {\xi}''{}-3 {\gamma}''{}\right)
\,
\Bigr]
+\frac{2 \varkappa ^2 {x^0}}{{x^3} } \left({\upsilon}{}+60 {\xi}{} \right)
$$
$$
\mbox{}
+24 
{\gamma}'{}
\left[
\,
2 {x^0} {\varphi}_{,3}{}-{x^3} \left(2 {x^0}^{{\alpha}+1} 
{\varphi}_{,22}{}-{\alpha} {\varphi}_{,1}{}+2 {x^0} {\varphi}_{,33}{}+2 
{x^0} {\varphi}_{,01}{}\right)
\,
\right]
$$
\be
\mbox{}
-16 \varkappa ^2 
{\xi}'{}
\left[
2 {x^0} {\varphi}_{,3}{}-{x^3} \left({x^0}^{{\alpha}+1} {\varphi}_{,22}{}
-{\alpha} {\varphi}_{,1}{}+{x^0} {\varphi}_{,33}{}+{x^0} {\varphi}_{,01}{}
\right)
\right]
=0
,
\label{B2EqField01}
\ee
$$
\mathbf{E_{02}}=
12 \left({\alpha} {\varphi}_{,2}{}+2 {x^0} {\varphi}_{,02}{}\right) 
{\gamma}'{}-4 \varkappa ^2 \left({\alpha} {\varphi}_{,2}{}+2 {x^0} 
{\varphi}_{,02}{}\right) {\xi}'{}
$$
\be
-{x^0} {\varphi}_{,2}{} {\varphi}_{,0}{} \left(\varkappa ^2 \left(8 {\xi}''{}+{\omega}\right)
-24 {\gamma}''{}\right)
=0
,
\label{B2EqField02}
\ee
$$
\mathbf{E_{03}}=
2 \varkappa ^2 {x^3} {\varphi}_{,3}{} \left({\alpha} ({\alpha}+2) {x^3}^2 {\varphi}_{,1}{}-4 {x^0}^2 
{\varphi}_{,0}{}\right) {\xi}''{}
$$
$$
\mbox{}
+2 \varkappa ^2 \left({\alpha} ({\alpha}+2) {x^3}^3 
{\varphi}_{,13}{}+{\alpha} ({\alpha}+2) {x^3}^2 {\varphi}_{,1}{}-4 {x^0}^2 
\left({\varphi}_{,0}{}+{x^3} {\varphi}_{,03}{}\right)\right) 
{\xi}'{}
$$
\be
\mbox{}
+{x^0}^2 \left({x^3} {\varphi}_{,3}{} {\varphi}_{,0}{} 
\left(24 {\gamma}''{}-{\omega} \varkappa ^2\right)+24 \left({\varphi}_{,0}{}+{x^3} 
{\varphi}_{,03}{}\right) {\gamma}'{}\right)
=0
,
\label{B2EqField03}
\ee
\be
\mathbf{E_{11}}=
8 {\varphi}_{,11}{} \left(3 {\gamma}'{}-\varkappa ^2 
{\xi}'{}\right)-{\varphi}_{,1}{}^2 \left(\varkappa ^2 \left(8 
{\xi}''{}+{\omega}\right)-24 {\gamma}''{}\right)
=0
,
\label{B2EqField11}
\ee
\be
\mathbf{E_{12}}=
8 {\varphi}_{,12}{} \left(3 {\gamma}'{}-\varkappa ^2 
{\xi}'{}\right)- {\varphi}_{,1}{} {\varphi}_{,2}{}\left(\varkappa ^2 
\left(8 {\xi}''{}+{\omega}\right)-24 {\gamma}''{}\right)
=0
,
\label{B2EqField12}
\ee
$$
\mathbf{E_{13}}=
24 \left({\varphi}_{,1}{}+{x^3} {\varphi}_{,13}{}\right) {\gamma}'{}
-{x^3} 
\Bigl[
{\varphi}_{,1}{}  {\varphi}_{,3}{} 
\left(\varkappa ^2 \left(8 {\xi}''{}+{\omega}\right)-24 {\gamma}''{}\right)
$$
\be
+8 \varkappa ^2 {\varphi}_{,13}{} {\xi}'{}
\Bigr]
=0
,
\label{B2EqField13}
\ee
$$
\mathbf{E_{22}}=
\mbox{}
-{x^3} 
\Bigl[
-{x^0} {x^3} 
\Bigl(
-{\omega} \varkappa ^2 {x^0}^{\alpha} 
{\varphi}_{,2}{}^2+{\varphi}_{,3}{}^2 \left(\varkappa ^2 \left(
16 {\xi}''{}+{\omega}\right)-48 {\gamma}''{}\right)
$$
$$
\mbox{}
+2  {\varphi}_{,0}{}{\varphi}_{,1}{} \left(\varkappa ^2 \left(16 {\xi}''{}+{\omega}\right)-48 {\gamma}''{}\right)
\Bigr)
$$
$$
\mbox{}
+8 \varkappa ^2 \left(4 {x^0} {\varphi}_{,3}{}-{x^3} \left({\alpha} {\varphi}_{,1}{}
+2 {x^0} {\varphi}_{,33}{}+4 {x^0} {\varphi}_{,01}{}\right)\right) {\xi}'{}
$$
$$
\mbox{}
-48 {x^0} \left({\varphi}_{,3}{}-{x^3} \left({\varphi}_{,33}{}+2 {\varphi}_{,01}{}\right)\right) {\gamma}'{}
\,
\Bigr]
$$
\be
\mbox{}
+2 \varkappa ^2 {x^0} {\upsilon}{}+120 \varkappa ^2 {x^0} {\xi}{}
=0
,
\label{B2EqField22}
\ee
$$
\mathbf{E_{23}}=
\mbox{}
-{x^3} {\varphi}_{,2}{} {\varphi}_{,3}{}  \left(\varkappa ^2 \left(8 
{\xi}''{}+{\omega}\right)-24 {\gamma}''{}\right)+24 
\left({\varphi}_{,2}{}+{x^3} {\varphi}_{,23}{}\right) {\gamma}'{}
$$
\be
\mbox{}
+8 \varkappa ^2 \left({\varphi}_{,2}{}-{x^3} {\varphi}_{,23}{}\right) {\xi}'{}
=0
,
\label{B2EqField23}
\ee
$$
\mathbf{E_{33}}=
\mbox{}
-\frac{2 {\alpha}^2 {x^3}^2 {\varphi}_{,1}{}^2 {\xi}''{}}{{x^0}^2}
-\frac{2 {\alpha}^2 {x^3}^2 {\varphi}_{,11}{} {\xi}'{}}{{x^0}^2}
+\frac{1}{2} {\omega} {x^0}^{\alpha} {\varphi}_{,2}{}^2
$$
$$
\mbox{}
-\frac{3 }{\varkappa ^2 {x^0} {x^3}^2}
\,
\Bigl[
\,
8 {x^0}  {x^3}^2 \left({x^0}^{\alpha} {\varphi}_{,2}{}^2
+2 {\varphi}_{,1}{} {\varphi}_{,0}{}\right) {\gamma}''{}
$$
$$
\mbox{}
+4 {x^3} \left({x^3} \left(2 {x^0}^{{\alpha}+1} {\varphi}_{,22}{}-{\alpha} {\varphi}_{,1}{}
+4 {x^0} {\varphi}_{,01}{}\right)-6 {x^0} {\varphi}_{,3}{}\right)
{\gamma}'{}
\Bigr]
$$
$$
\mbox{}
+8 {x^0}^{\alpha} {\varphi}_{,22}{} {\xi}'{}
+8 {x^0}^{\alpha} {\varphi}_{,2}{}^2 {\xi}''{}
-\frac{4 {\alpha} {x^3}^2 {\varphi}_{,1}{}^2 {\xi}''{}}{{x^0}^2}
-\frac{4 {\alpha} {x^3}^2 {\varphi}_{,11}{} {\xi}'{}}{{x^0}^2}
-\frac{4 {\alpha} {\varphi}_{,1}{} {\xi}'{}}{{x^0}}
$$
\be
\mbox{}
-\frac{1}{2} {\omega} {\varphi}_{,3}{}^2+{\omega} {\varphi}_{,0}{} {\varphi}_{,1}{}
+16  {\varphi}_{,0}{} {\varphi}_{,1}{} {\xi}''{}
+16 {\varphi}_{,01}{} {\xi}'{}
+\frac{{\upsilon}{}}{{x^3}^2}
+\frac{60 {\xi}{}}{{x^3}^2}
=0
,
\label{B2EqField33}
\ee
where we used the replacement $V={\upsilon}(\varphi)-3/\varkappa^2$.

The scalar field equation (\ref{EqScalar}) for the metric (\ref{MetricB2}) becomes:
$$
\frac{{x^3}}{2 {x^0}}\,
 \left[
 {x^3} \left(2 {x^0}^{{\alpha}+1} {\varphi}_{,22}{}-{\alpha} {\varphi}_{,1}{}
+2 {x^0} {\varphi}_{,33}{}+4 {x^0} {\varphi}_{,01}{}\right)
-4 {x^0} {\varphi}_{,3}{}
\right]
$$
\be
\mbox{}
+
\frac{1}{{\omega}}
\,
\left(
\frac{72 {\gamma}'{}}{\varkappa ^2}-{V}'{}-84 {\xi}'{}
\right)
=0
,
\label{B2EqScalar}
\ee
We will look for a solution of the field equations in the class of scalar fields that have a ''split'' form:
\be
\varphi(x^0,x^1,x^2,x^3)=\varphi_0(x^0)\,\varphi_1(x^1)\,\varphi_2(x^2)\,\varphi_3(x^3)
.
\ee
The compatibility of the field equations leads to the condition:
\be
\varphi_{,1}=0
\quad \to \quad
\varphi=\varphi(x^0,x^2,x^3).
\ee
The variable $ x^1 $ is an ignored isotropic variable that the metric does not depend on.
Thus, the scalar field $\varphi$ does not depend on the ignored isotropic variable~{$ x^1 $}.

\subsection{Exact solution for model II-B2 in case $\varphi_{,2}\ne 0$.}

Consider the case where the scalar field depends on the ignored variable $ x^2 $. Then the compatibility conditions for the equations (\ref{B2EqField02}),  (\ref{B2EqField03}) and (\ref{B2EqField23})
  lead to the ratios:
\be
0=
{\alpha} {\varkappa} ^2 {\varphi_0}{} {\varphi_3}{} {\varphi_2}'{} {\varphi}
   {\xi}'{} \left(3 {\gamma}'{}-{\varkappa} ^2
   {\xi}'{}\right)
   ,
\ee
\be
0=
{\varkappa} ^2 {\varphi_3}{} {\varphi_0}'{} {\varphi_2}'{} {\varphi}
   {\xi}'{}
,
\ee
\be
0=
{\varphi_2}'{} {\varphi} \left({\alpha} {x^3} {\varphi_0}{} {\varphi_3}'{} \left(3
   {\gamma}'{}-{\varkappa} ^2
   {\xi}'{}\right)-6 {x^0} {\varphi_3}{} {\varphi_0}'{}
   {\gamma}'{}\right)
.
\ee
Hence follows the condition:
\be
\xi(\varphi)=\mbox{const}.
\ee
Thus, the considered spatially homogeneous wave-like model of space-time does not admit the presence of a dynamic term associated with the Gauss-Bonnet invariant.

In addition, the compatibility conditions lead to the equations:
\be
{\alpha} {x^3} {\varphi_3}'{} / {\varphi_3}{}={\sigma},
\qquad
2 {x^0} {\varphi_0}'{} /{\varphi_0}{}
={\sigma},
\qquad
{\sigma} - \mbox{const}.
\ee
Thus, the scalar field takes the form
\be
\varphi(x^0,x^2,x^3) ={ x^0}^{{\sigma}/2}{x^3}^{{\sigma}/\alpha}\varphi_2(x^2),
\ee
Then the equation (\ref{B2EqField23}) gives a second order differential equation on $ \gamma (\varphi) $, the solution of which has the form
\be
\gamma(\varphi)=
c_1
\,
{\varphi}^{-{\alpha}/{\sigma}}
+c_2+\frac{{\omega}
   {\sigma} {\varkappa} ^2 }{48 ({\alpha}+2 {\sigma})} \, {\varphi}^2
,\qquad
c_1,c_2 - \mbox{const}
.
\ee
In this case, the scalar equation (\ref{B2EqScalar}) takes the following form:
\be
{\varphi} \left[
{\varphi}^{2\alpha/{\sigma}}
{\varphi_2}^{-1-2\alpha/{\sigma}} {\varphi_2}''{}
 +\frac{{\sigma} ({\sigma}-3 {\alpha}) }{\alpha^2}
\right]
=
\frac{{\upsilon}'({\varphi})}{{\omega}}
-\frac{72 {c_1} {\varphi}^{-\frac{{\alpha}+{\sigma}}{\sigma}}}{{\omega} {\varkappa} ^2}
-\frac{3 {\sigma} {\varphi}}{{\alpha}+2 {\sigma}}
,
\ee
this implies
\be
{\varphi_2}^{-1-2\alpha/{\sigma}} {\varphi_2}''{}=\mbox{const}
\quad \to \quad
{\varphi_2}'=\rho\,{\varphi_2}^{1+\alpha/{\sigma}} 
,\quad
\rho- \mbox{const}
.
\ee
Then the scalar equation (\ref{B2EqScalar}) takes the following form:
\be
 \frac{{\alpha}^2 {\rho}^2  ({\alpha}+{\sigma})  
{\varphi}^{2\alpha/{\sigma}}
+{\sigma}^2 ({\sigma}-3 {\alpha}) 
}{{\alpha}^2 {\sigma}}
\,
{\varphi}
=
\frac{{\upsilon}'({\varphi})}{{\omega}}
-\frac{72 {c_1} {\varphi}^{-\frac{{\alpha}+{\sigma}}{\sigma}}}{{\omega} {\varkappa} ^2}
-\frac{3 {\sigma} {\varphi}}{{\alpha}+2 {\sigma}}
.
\ee
From here we can find the potential of the scalar field 
\be
V(\varphi ) =
 \frac{{\omega} {\rho}^2}{2}\,
 {\varphi}^{\frac{2 ({\alpha}+{\sigma})}{\sigma}}
 -\frac{72 {c_1} {\sigma} {\varphi}^{-\frac{\alpha}{\sigma}}}{{\alpha} {\varkappa} ^2}
 -\frac{{\omega} {\sigma}^2 (5 {\alpha}-2 {\sigma}) {\varphi}^2}{2 {\alpha}^2 ({\alpha}+2 {\sigma})}
 +{c_3}
.
\ee
Compatibility conditions for field equations (\ref{B2EqField01}), (\ref{B2EqField22}) and (\ref{B2EqField33}) lead to the following form of integration constants:
\be
c_1=0
,\qquad
c_2=(1+8 {\varkappa}^2 \xi)/24
,\qquad
c_3=-60 \xi.
\ee
Then from the equation (\ref{B2EqField00}) we obtain the value of the constant ${\sigma}$:
\be
{\sigma}={\alpha}/2
.
\ee
Finally, for the case $ \varphi_{, 2} \ne 0 $, we obtain the following exact solution of the field equations (\ref{EqsField}) and (\ref{EqScalar}) in vacuum for the metric (\ref{MetricB2}):
$$
ds^2= \frac{1}{{x^3}^2}\left(
2\,dx^0dx^2
+{x^0}^{-{\alpha}}\,{dx^2}^2+{dx^3}^2
\right),
$$
\be
\xi(\varphi ) = {\xi}=\mbox{const}
,\qquad
\gamma(\varphi ) = \frac{{\omega} {\varkappa}^2}{192} \, {\varphi}^2+\frac{1}{24} \left(8 {\xi} {\varkappa} ^2+1\right)
,
\ee
\be
V(\varphi ) = \frac{{\omega} }{8\,{a}^4} \,{\varphi}^6-\frac{ {\omega}}{4} \, {\varphi}^2-60 {\xi}-\frac{3}{{\varkappa} ^2}
,
\ee
\be
\varphi (x^0,x^2,x^3)= 
{a} \left( {x^0} \right)^{{\alpha}/4} \sqrt{\frac{x^3}{x^2-{b}}},
\qquad
{a},{b} - \mbox{const}
.
\ee
Here $ \alpha $ is a constant parameter of the model, the constant $ \Lambda = -3 \, (20 \, \xi + 1 / {\varkappa}^2) $
plays the role of a cosmological constant.
The model becomes conformally flat (but not flat) when $ \alpha (\alpha + 2) = 0 $.
The variable $ x^0 $ has a wave character (along  $ x^0 $ the space-time interval vanishes).

Thus, in a spatially homogeneous wave-like model of Shapovalov's space-time of type II-B2, when the scalar field depends on the ''ignored'' variable $ x^2 $, the scalar potential $ V (\varphi) $ and the function $ \gamma (\varphi) $ 
 in the Lagrangian depend on the scalar field polynomially with even degrees. The term associated with the Gauss-Bonnet invariant contributes only to the cosmological constant.
 
\subsection{Exact solution for the model subtype II-B2 in the case when the scalar field depends only on non-ignored variables ($\varphi_{,1}=\varphi_{,2}= 0$).}

Consider the solution of the field equations (\ref{EqsField}), (\ref{EqScalar}) for the metric (\ref{MetricB2}) in the case when the scalar field depends only on non-ignored variables (on which the metric depends):
\be
\varphi=\varphi(x^0,x^3)
.
\ee
Equation (\ref{B2EqField33}) implies that
\be
\varphi_{,3}={F}(\varphi)/x^3
,
\label{B2phi3}
\ee
\be
\gamma'(\varphi)=
\varkappa ^2
\,
\frac{ 
{\omega} {{F}}(\varphi )^2-2 \left[{\upsilon}(\varphi )+60 {\xi}(\varphi )\right]
}{144 {{F}}(\varphi)}
   ,
\label{functiongamma}
\ee
where ${F}(\varphi)$ is auxiliary function of a scalar field.

Then the equation for the scalar field (\ref{EqScalar}) takes the form:
\be
{{\omega}}
{{F}}{} \left({{F}}'{}-3\right)=
\frac{
-{\omega} {{F}}{}^2+2 {{F}}{} \left({\upsilon}'{}
+84 {\xi}'{}\right)+2 ({\upsilon}{}+60 {\xi}{})
}{2 {{F}}{}}
.
\ee

From the scalar equation we obtain
\be
{{F}}'(\varphi)=
\frac{5 {\omega} {{F}}{}^2+2 {{F}}{}
   \left({\upsilon}'{}+84 {\xi}'{}\right)+2
   ({\upsilon}{}+60 {\xi}{})}{2 {\omega}
   {{F}}{}^2}
   .
\label{functionV}
\ee
Тогда условия совместности уравнений (\ref{B2EqField02}),
 (\ref{B2EqField03}),  (\ref{B2EqField23})
 приводят к требованию:
 \be
 \xi=\mbox{const}.
 \ee
Thus, the considered spatially homogeneous wave-like model of space-time does not allow the presence of a ''dynamic'' term associated with the Gauss-Bonnet invariant, which contributes only to the cosmological constant.

From the system of field equations, only the equation (\ref{B2EqField00}) remains, which takes the form
$$
0=
 {\omega}{\alpha} ({\alpha}+2)
\Bigl(
\,
72    {\gamma}{}
-3  \left(8 {\xi} \varkappa ^2+1\right)
   \,
\Bigr)
    {{F}}{}^4
$$
$$
\mbox{}
   +{x^0}^2 
\Bigl[
\,
{\varphi}_{,00}
\Bigl(
2   {\omega}^2 \varkappa ^2 {{F}}{}^5
-4  {\omega} \varkappa ^2 {{F}}{}^3  ({\upsilon}{}+60 {\xi})
\Bigr)
$$
\be
\mbox{}   
+
{\varphi}_{,0}{}^2
\Bigl(
4 \varkappa^2 ({\upsilon}{}+60 {\xi})
\left(
3 {\omega}  {{F}}{}^2
+ {{F}}{}    {\upsilon}'{}
+ {\upsilon}{}+60 {\xi}
\right)
 -7   {\omega}^2 \varkappa ^2  {{F}}{}^4
-2   {\omega} \varkappa ^2 {{F}}{}^3  {\upsilon}'{}
\Bigr)
   \,
\Bigr]
   ,
\ee

We will look for a solution for a scalar field in a ''split'' form:
\be
\varphi(x^0,x^3)=\varphi_0(x^0)\varphi_3(x^3)
,
\ee
then from the relation (\ref{B2phi3}) we obtain
\be
x^3{\varphi_3}'/{\varphi_3}=\mbox{const}={\sigma}
\quad
\to
\quad
{\varphi_3}={x^3}^{\sigma}
,\qquad
{{F}}(\varphi)={\sigma} \varphi
.
\ee
Integrating the equation (\ref{functionV}), we get
\be
V(\varphi ) = {c_1} {\varphi}{}^{-1/{\sigma}}
+\frac{{\omega} {\sigma}^2 (2 {\sigma}-5) }{4 {\sigma}+2} {\varphi}{}^2
-60 {\xi}-\frac{3}{\varkappa^2}
,\qquad
{\sigma}\ne -1/2
.
\ee
Then from the equation (\ref{functiongamma}) we have
\be
\gamma(\varphi)=
{ c_2}+
\frac{\varkappa ^2 \varphi^{-1/{\sigma}} 
\left[
{c_1} (4 {\sigma}+2)+3 {\omega} {\sigma}
   \varphi^{2+{1}/{\sigma}}
\right]
}{144 (2 {\sigma}+1)}
 .
\ee
Equation (\ref{B2EqField00}) becomes (${\sigma}\ne-1/2$):
$$
0=
\frac{
{c_1} {\varphi}{}^{-1/{\sigma}} 
\left(
{\alpha} ({\alpha}+2) {\sigma}^2 {\varphi_0}{}^2{x^0}^{-2}
-4 {\sigma}  {\varphi_0}{} {\varphi_0}''{}
+4 ({\sigma}+1)  {\varphi_0}'{}^2
\right)
}{12 {\sigma}^2 {\varphi_0}{}^2}
$$
$$
\mbox{}
+
\frac{
{\omega} {\varphi}{}^2 
\left(
{\alpha} ({\alpha}+2) {\sigma} {\varphi_0}{}^2{x^0}^{-2}
+8 {\sigma}  {\varphi_0}{} {\varphi_0}''{}
-8 ({\sigma}+1)  {\varphi_0}'{}^2
\right)
}{8 (2 {\sigma}+1) {\varphi_0}{}^2}
$$
\be
\mbox{}
-\frac{{\alpha} ({\alpha}+2) \left(-24 {c_2}+8 {\xi} \varkappa ^2+1\right)}{4 \varkappa ^2 {x^0}^2}
.
\label{B2EqPhi0}
\ee
From the equation (\ref{B2EqPhi0}) we obtain the following consequences:
\be
\alpha(\alpha+2)
\left(
{c_2}-(1+8 {\xi} \varkappa ^2 )/24
\right)=0
,
\ee
\be
\alpha(\alpha+2){c_1}=0
,
\ee
\be
{\alpha} ({\alpha}+2) {\sigma} {\varphi_0}{}^2{x^0}^{-2}
+8 {\sigma}  {\varphi_0}{} {\varphi_0}''{}
-8 ({\sigma}+1)  {\varphi_0}'{}^2=0
,
\ee
If $ {\alpha} ({\alpha} +2) = 0 $, then the space-time degenerates -- becomes conformally flat.

For completeness, we present the exact solution of the field equations
for this ''degenerate'' case of conformally flat spacetime:
$$
ds^2= \frac{1}{{x^3}^2}\left(
2\,dx^0dx^2
+{x^0}^{-{\alpha}}\,{dx^2}^2+{dx^3}^2
\right)
,\qquad
\alpha(\alpha+2)= 0
,
$$
\be
\varphi=
\varphi(x^0,x^3)=
c_1 
\left(
\frac{x^3
}{
c_2{x^0}+c_3
}
\right)^{\sigma}
,
\ee
\be
\gamma(\varphi ) = \frac{\varkappa ^2 
 \left(
 3 {\omega} {\sigma} {\varphi}{}^{2}
 +2{c_4} (2 {\sigma}+1) {\varphi}{}^{-1/{\sigma}}
\right)
}{144 (2 {\sigma}+1)}+{c_5}
,
\ee
\be
V(\varphi ) = {c_4} {\varphi}{}^{-1/{\sigma}}
+\frac{{\omega} {\sigma}^2 (2 {\sigma}-5) }{2(2 {\sigma}+1)}
\,{\varphi}^2
-60 {\xi}-\frac{3}{\varkappa ^2}
.
\ee
$$
\xi(\varphi ) = {\xi} - \mbox{const},
\qquad
{\sigma}, c_1,c_2,c_3,c_4,c_5  - \mbox{const},
\qquad
{\sigma}\ne 0, -1/2.
$$

Let us now consider the main case when the space-time is not conformally flat, i.e. $ \alpha (\alpha + 2) \ ne 0 $.

Thus, the final solution of the field equations for the spatially homogeneous wave-like Shapovalov model of type II-B2
$$
ds^2= \frac{1}{{x^3}^2}\left(
2\,dx^0dx^2
+{x^0}^{-{\alpha}}\,{dx^2}^2+{dx^3}^2
\right)
,\qquad
\alpha(\alpha+2)\ne 0
$$
for the quadratic theory of gravity with a scalar field takes the form:
\be
\varphi(x^0,x^3)=
c_1
\left(
x^3\,
\frac{
{x^0}^{-1/2+\sqrt{2+{ {\alpha} ({\alpha}+2)}/{\sigma} }}
}{
c_2+{x^0}^{\sqrt{1+{{\alpha} ({\alpha}+2)}/({2 {\sigma}})  }}
}
\right)^{\sigma}
,\quad
\alpha,
\sigma,
c_1, c_2  - \mbox{const},
\ee
$$
{\sigma}\ne 0,  -1/2;
\qquad
\xi(\varphi ) = {\xi} - \mbox{const}
,
$$
$$
\gamma(\varphi ) = \frac{
 \varkappa ^2  {\omega} {\sigma} 
}{48 (2 {\sigma}+1)}
\,{\varphi}^{2}
+(1+8 {\xi} \varkappa ^2 )/24
,
$$
$$
V(\varphi ) = \frac{{\omega} {\sigma}^2 (2 {\sigma}-5) }{2(2 {\sigma}+1)}
\,{\varphi}^2
-60 {\xi}-\frac{3}{\varkappa ^2}
.
$$
The constant $ \Lambda = 3 (20 {\xi} + {1} / {\varkappa ^ 2}) $ plays the role of the cosmological constant.

Thus, in a spatially homogeneous wave-like model of Shapovalov's space-time of type II-B2, when the scalar field depends only on non-ignored variables, the scalar potential $ V (\varphi) $ and the coefficient $ \gamma (\varphi) $ at $ R^2 $ in the Lagrangian depend on the scalar field quadratically. The term associated with the Gauss-Bonnet invariant contributes to the cosmological constant, while does not contribute to the ''scalar'' dynamics of the model.

\section*{Conclusion}

In this work, a number of exact solutions are obtained for spatially homogeneous wave-like models of Shapovalov space-time of type II in the theory of gravity quadratic in curvature with a scalar field.
The form of the scalar field, scalar potential, and scalar field dependent functions included in the Lagrangian of the theory is found. In the considered models, the term in the Lagrangian associated with the Gauss-Bonnet invariant does not contribute to the ''scalar'' dynamics, but only contributes to the cosmological constant.
The obtained exact models can be interpreted as models of gravitational wave disturbances at the early stages of the development of the Universe. The considered models make it possible to integrate in quadratures the equations of motion of test particles in the Hamilton-Jacobi formalism. The solutions obtained provide, among other things, precisely integrable models for debugging numerical methods of analysis in modified $F (R, G)$ - theories of gravity. 

\section*{Acknowledgments}

The reported study was funded by RFBR, project number N~20-01-00389~A.

%

\end{document}